\documentstyle[11pt,psfig]{article}
\newcommand{\eqn}[1]{eqn.~(\ref{eq:#1})}
\newcommand{\fig}[1]{fig.~\ref{fig:#1}}
\newcommand{\se}[1]{section~\ref{sec:#1}}

\newcommand{\vv}{\vec{v}}
\newcommand{\tm}{t_{\rm m}}
\newcommand{\la}{\langle}
\newcommand{\ra}{\rangle}

\title{Simulation of forces on polymers due to slippage}
\author{J.M. Deutsch \quad Hyoungsoo Yoon \\
        University of California, Santa Cruz \\
        Santa Cruz, CA 95064}
\begin{document}
\maketitle

\begin{abstract}
We consider the force on the end of a polymer chain being
pulled through a network at velocity $v$, using computer simulations.
We develop algorithms for measuring the force on
the end of the chain using
lattice models of polymers. Our algorithm attaches
a spring to the end being pulled and uses its average
extension to calculate the force. General problems associated
with the use of lattice models in obtaining forces
are discussed. Variants of this method are used to 
obtain upper and lower bounds to the force.
The results obtained are in agreement with recent
analytical predictions and experiments.
\end{abstract}

\section{Introduction}
Grafted polymer chains have been used to alter 
physical properties of interfaces. 
Colloid particles are maintained in suspension
by the addition of a dense layer of end-tethered
chains.  Mixing of incompatible polymeric materials is
facilitated by using diblock copolymers which
stabilize the interface of microdomains. 
Reinforcement of junctions by grafting connector
chains is yet another example of the numerous
applications of grafted polymers.

In this paper we will focus on the force
of a chain pulled through a network. This
is closely related to the problem of
how to reduce slippage of a viscous
polymer melt. 
De~Gennes\cite{PGdG:ConcentratedPolymer}
proposed that
this could be done by the addition
of grafted chains. 
This was considered further in the case of
a chain being pulled though a network by
Rubinstein et al.\cite{MRAAetal:SlippageRubber},
and later generalized to a melt%
\cite{AAFBWetal:SlippageEntangled},
a network being the limit of a melt with
extremely long chains.
It was proposed that the efficiency of slippage
reduction by grafted chains is not as great
as one might first expect because of the
non-homogeneous deformation of these chains%
\cite{FBPGdG:ShearDependent}.
Because our study will focus on the force
of a polymer chain being pulled through
a network we will restrict out discussion
to this case.

Rubinstein et al.\cite{MRAAetal:SlippageRubber}
considered the case of a single
polymer chain being pulled at one end at a 
constant velocity $v$. They pointed out that
the mode of relaxation for the chain is similar
to the arm of a star polymer%
\cite{PGdG:ReptationStars}.

Using scaling arguments, they predicted three
separate regimes of behavior of the chain as
a function of the pulling velocity $v$ and the
chain length $L$. In the first regime, that
of low speeds, they predicted that the frictional
force $f$ is proportional to $v$. The coefficient
of proportionality should scale inversely with 
the longest relaxation time for the chain 
$T_{\rm rel} \propto \exp({\rm const.}\times L)$. 
Throughout
this work we will take the gel spacing to be
one.

When the velocity reaches a certain value, the
chain can no longer completely equilibrate before
it is pulled further forward. At this point the
part of chain closest to the point being pulled
denoted the head, adopts a stick-like conformation
while the tail region of the chain can still
adopt a gaussian conformation. Thus one has
the picture of the chain being stick-like
at the front and then forming a plume in the
back. As the velocity increases, so does the
length of the stick region, and the plume
size diminishes.
In this regime one expects the force
to be almost independent of velocity
as the tension in the chain is almost at
its equilibrium value, but now it is directed
opposite the direction of $\vv$. The
tension changes little because these pulling
speeds are still very low in comparison
with speeds related to the relaxation time
of a Rouse chain, that is $v \sim 1/L^2$.

The third regime, that of high velocity,
the plume completely disappears and the
frictional force increases linearly with
pulling speed. The sole contribution to
the force comes from the Rouse friction of
the constituent monomers.

Some recent experimental data appears to agree with 
these theoretical predictions%
\cite{HRB:ChainPullout,KMHHetal:slippage}.
In this paper we present further evidence for this
behavior using numerical simulations.

Simulation measurement of
forces in general, pose interesting questions in computer
simulations. Since the value of the forces in the present
situation are quite small, long runs are necessary to recover
a high signal to noise ratio. Because of this fast
algorithms such as lattice Monte Carlo
models are highly desirable. However it is not obvious how
to extract a force out of a lattice simulation. We have
devised some new methods for measuring forces which should
generalize well to other problems involving polymers.

In the following section, we describe the models used
and the simulation procedure. We will explain how and where
the methods used to measure forces fail and what regimes
one expects the results to be valid. In \se{results} we
present our numerical results. In \se{conclusion} we
present our conclusions.

\section{The Models}
\label{sec:models}
\subsection{The Chain}
\label{sec:chain}

\begin{figure}[p]
\begin{center}
\ 
\psfig{file=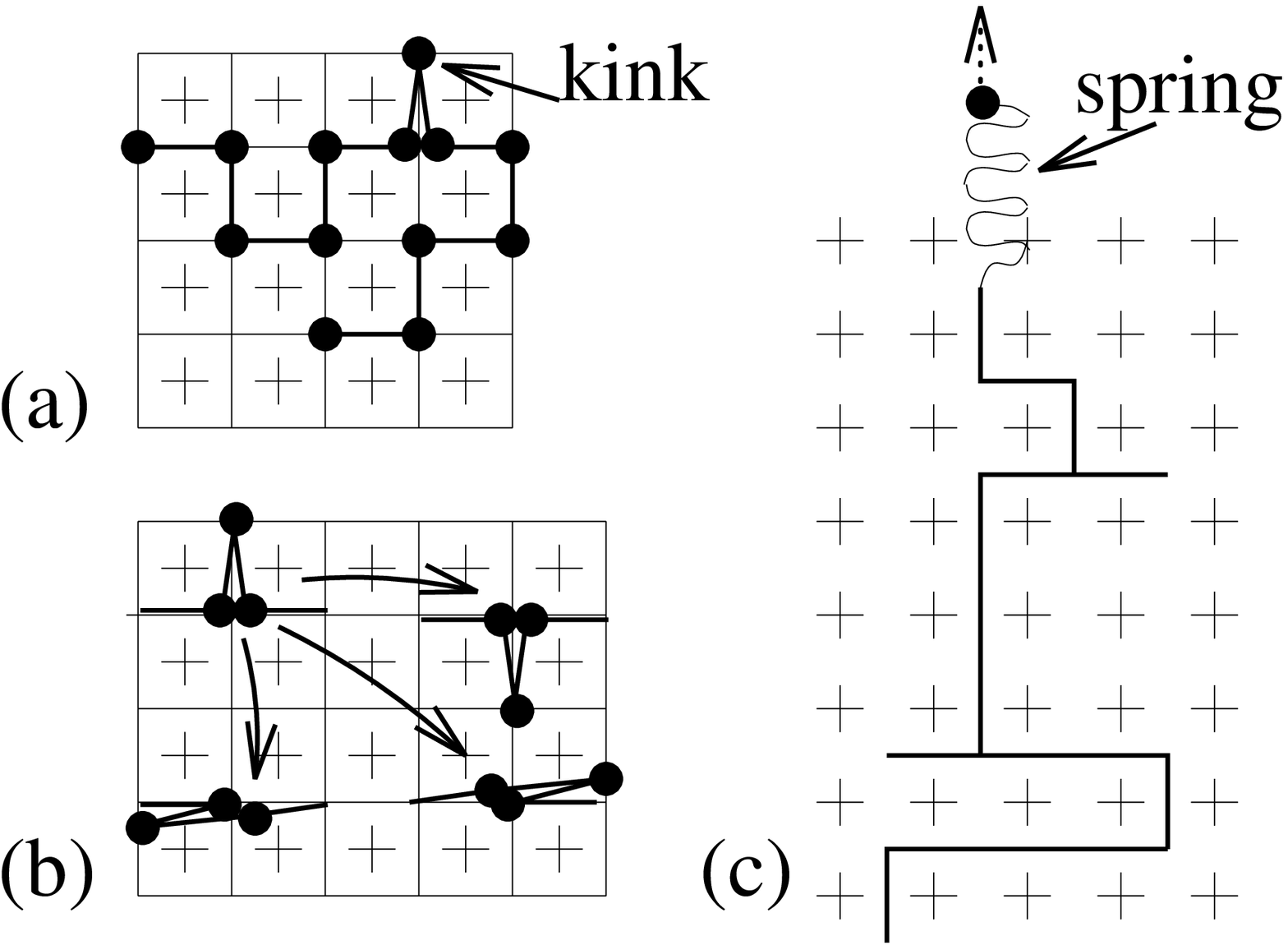,width=6in}
\end{center}
\caption[]{Illustration of the lattice model that was used
in the simulations. (a) The links of the chain are represented 
by the thick black lines and the monomers by the filled 
circles. The +'s represent entanglements that cannot be crossed.
(b) The elementary moves allowed in this simulation. The kink
on the upper left hand corner can move to one of the four 
conformations shown with equal probability. (c) The force is
measured by attaching a spring to the head of the chain and
pulling the other end of the spring, denoted by the filled
circle, at velocity $v$.}
\label{fig:cage}
\end{figure}

We consider a chain on a two dimensional square lattice of 
length~$L$. The distance between monomers, or beads, is
one lattice spacing. The fact that the chain is in a network
is represented by entanglement points, or a ``cage'' that
lives between lattice sites at the center of squares formed
by adjacent lattice sites, as shown in \fig{cage} by
the +'s. In our simulation
we only considered the case where the cage spacing is one.
We neglect the effects of excluded
volume and allow the chain to move following two sets of
dynamics. 

The first allows only for short range moves
and is called the Evans Edwards model%
\cite{KEESFE:ComputerSimulation:1,%
SFEKEE:ComputerSimulation:2,%
KEESFE:ComputerSimulation:3}.
The backbone of
the chain is not allowed to cross entanglement points, 
and therefore the only moves that can take place are that
of ``kinks''. Kinks are local conformations of two links
that are doubled up on top of each other. The dynamics
proceed
as follows. A monomer is picked at random
from the chain. If it is not the tip of a kink,
no move is attempted. If it is,  then it is flipped
randomly with equal probability to any one of four conformations
as shown in \fig{cage}(b).
The tail of the polymer is handled in much the same way
as a kink. 

The head of the polymer, that is the point being pulled,
is modeled in a slightly more complicated manner. We consider
it being pulled by an anisotropic spring, with a spring constant 
for vertical motion of $k$. Motion of the end in the horizontal
direction is prohibited.
Therefore the end of the chain is not pulled directly 
but by a somewhat stiff spring shown in \fig{cage}(c). 
The other end of the spring represented in the figure by the solid circle
is pulled with velocity $v$. We will discuss below the precise
way that we move this spring. Motion of the head of the
chain then proceeds by the normal Metropolis algorithm.
The potential energy of the system is that of the spring
which is $k \Delta z^2 /2$ where $\Delta z$ is the spring
length.

The second algorithm we consider is one developed by one
of the authors previously to study electrophoresis in
strong fields\cite{JMDJDR:SimulationHighly}.
This long range algorithm
obeys detailed balance, and it was shown to more correctly
model the motion of chains under high tension than short
range models, which dramatically fail to describe 
electrophoresis\cite{MOdlCJMDetal:ElectrophoresisPolymer}.
It adds long range moves in addition to
the short range moves described above. In one step, 
a kink can move anywhere between its two adjacent kinks. 
The details of the algorithm and its implementation
are described in ref.~\cite{JMDJDR:SimulationHighly}.

\subsection{The Force}
\label{sec:force}

The force is measured in three ways. They are all based
on relating the length of the spring to the force on
the chain. If the system was in equilibrium,
for example if the chain was hanging from a spring
that was tethered to a fixed point, then it is straightforward
to relate the average displacement of the spring $\la\Delta z\ra$
to the force at the end of the chain. The energy
of the system is $E = k \Delta z^2 /2 - f\Delta z$.
Therefore at fixed temperature, taken to be unity,
$\la\Delta z\ra$ is a function of $f$
\begin{equation}
\la\Delta z\ra = {\sum_{\Delta z} \Delta z e^{-E} \over
\sum_{\Delta z}  e^{-E}}
\label{eq:spring}
\end{equation}
There is a measurement time $\tm$ necessary to
get a good estimate of the force. Notice
that the spring constant should not be chosen to be too
large. Otherwise $\tm$ becomes very long ${} \propto \exp(k/2)$.

The end of the spring jumps discretely between even
lattice sites. The reason we cannot simply move the
end to adjacent lattice sites is that the square
lattice is bipartite and as a consequence of
the dynamics described above, the end
of the chain can only live on half the sites.
The time spent at a given
lattice site is $\Delta t = 2/v$, recalling that
we have chosen the lattice spacing to be one.

One would expect that when the velocity is small
this formula should still apply as the measurement
time $\tm$ will be much less than $\Delta t$.
However when the velocity of the chain is sufficiently high
one expects that this formula should break down because
now the spring is always out of equilibrium and therefore
\eqn{spring} cannot apply.

We also note that we cannot make $k$ too small
or this would allow the end of the chain to
move far from the point it is being tethered.
We do not want this, as 
it would alter the physics of the problem considerably.

The method above, by Newton's third law should certainly be valid in a
continuum, when we consider moving the end of the spring continuously
at constant velocity $v$.  The definition of force on a lattice is an
extension of force in a continuum. However there are many possible ways
of extending its definition. We are interested in the definition that
best mimics this continuum behavior. The fact that on a lattice we are
forced to move the chain in a discontinuous manner creates some
additional difficulties we must be aware of.  These are
related to the fact that one needs $\tm \ll \Delta t$. 
As a consequence, the
above method for measuring force should result in an answer that is
expected to be an upper bound to the true continuum force.

To understand this, consider the extreme case where the spring
coefficient is very large. We expect that the answer for the
force $f$ should be almost independent of the value of $k$ chosen, and
therefore the average length of the spring $\la\Delta z\ra = f/k$ which
we are assuming is much less than $1$.
However when the end of the spring moves forwards by two lattice
spacings, this generates an instantaneous spring length of $2$ as
the chain has not yet had a chance to move. On a continuum this
would never happen as the chain is moved in infinitesimal steps
so that the spring length never deviates much from its small equilibrium
value. Here however this initial transient behavior can dominate the
results for large $k$, resulting in too high a measured force.
Even if $k$ is not very large, one would expect to notice some
effect due to this problem.

A way to avoid this problem is to neglect the initial transient
behavior. That is after moving the end of the spring, one only starts to
measure $\Delta z$ {\em after\/} the first time $\Delta z$ becomes zero.
This cuts off the initial transient behavior and should get rid of the
above problem. 
However it also neglects the average of $\Delta z$
in this time interval. 
This is not as serious a problem at low speeds,
as first it might seem. The average time it takes $\Delta z$
to become zero is short. It scales as the time it takes the nearest kink
to diffuse to the head of the chain. Because in most situations
we consider, the kink density remains of order unity, this time
is of order the time for motion of individual kinks. On the other hand,
the relaxation time for the tension of the chain scales as a power of chain
length. Therefore for long chains and kink densities of order
unity, we can safely cut off this initial transient behavior.
Because this initial transient is neglected
the result should give a lower bound to the continuum force. 

An intermediate strategy for obtaining the force that does not
cut off the initial transient behavior nor give too large
a value for $\la\Delta z\ra$ is as follows. We still keep
fixed time intervals with duration $\Delta t$
in this method, but now the beginning of an 
interval does not signify the spring 
end moving forward. Instead, at the start of a new interval
the spring end is moved only when the head
of the chain is {\em above it\/} by two lattice spacings. This insures that
when the spring end moves forward, its $\Delta z$ is zero.
We calculate the force by time averaging $\Delta z$ 
as in the first method.
If $v$ is sufficiently small, the chances of the head never
moving above the spring end is negligibly small, however at
high velocities this will break down, and in some time intervals
it will not move forwards. In this case the velocity is
not constant but will fluctuate. We will only consider velocities
that are sufficiently low so that this is not a problem.
A further problem with this method is that we have
introduced a ``Maxwell's demon'' by choosing exactly when to
move the head of the polymer forwards. Of order $k_B T$ of
free energy is expended in making this decision. This is 
expected to reduce the force somewhat over a short time-scale,
which again should be the microscopic kink-jump time. As argued
in the previous paragraph, this initial transient can be neglected
for long chains and kink densities of order unity.

In summary, we consider three different definitions for the
force obtained by measuring the length of a spring that is pulling 
the head of a chain and then relating it to the force 
through \eqn{spring}. First we use the time averaged spring length
which should lead to an upper bound to the continuum force.
Second we do not time average the spring length when the
end of the spring is first moved, but wait until the first
time where the spring length is zero. Third we average
over all times but only move the spring end when the head of
the chain is above it, to insure that the length of the spring
is initially zero.

\section{Results}
\label{sec:results}

\begin{figure}[p]
\begin{center}
\ 
\psfig{file=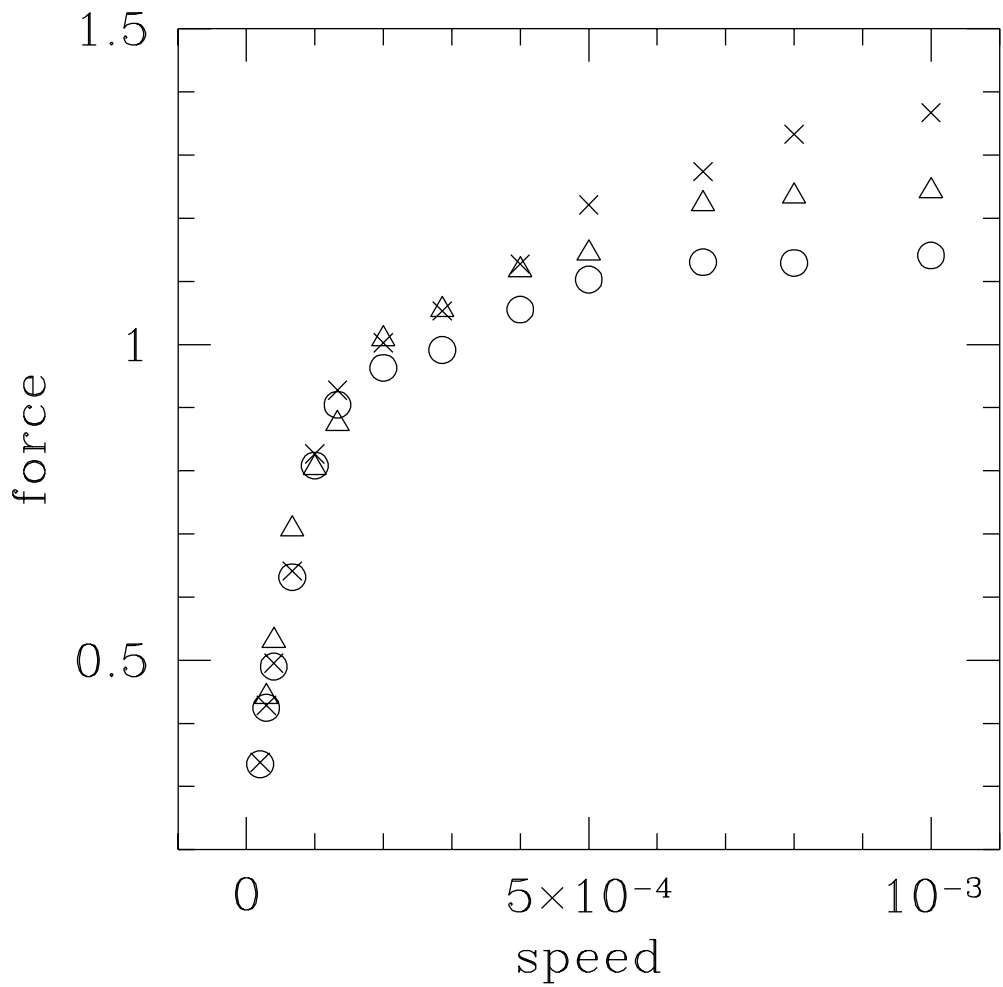,width=6in}
\end{center}
\caption[]{The force versus velocity for chains of length 20
using the short range moves. The crosses show the first the
method of measuring the force, the circles represent the second,
and the triangles represent the third.}
\label{fig:force}
\end{figure}

We now present the results of our simulations. Measurements
for each data point were done over a period where
the chain went $20$ times its total length. We chose a spring
constant $k = 1$. Smaller values were also tested and we found
that our results differed only slightly. The force as a function
of velocity is shown in \fig{force} for a chain of length $20$ for the
case of short range moves. The different symbols
show the three different methods for measuring the force. The first
method is shown with the crosses, the circles represent the second
method and the triangles represent the third method. Note that
for low speeds, all three methods give the same results within
statistical error, but at high speeds they deviate somewhat.
We expect that the crosses represent an upper bound and the
circles a lower bound to the continuum force.

\begin{figure}[p]
\begin{center}
\ 
\psfig{file=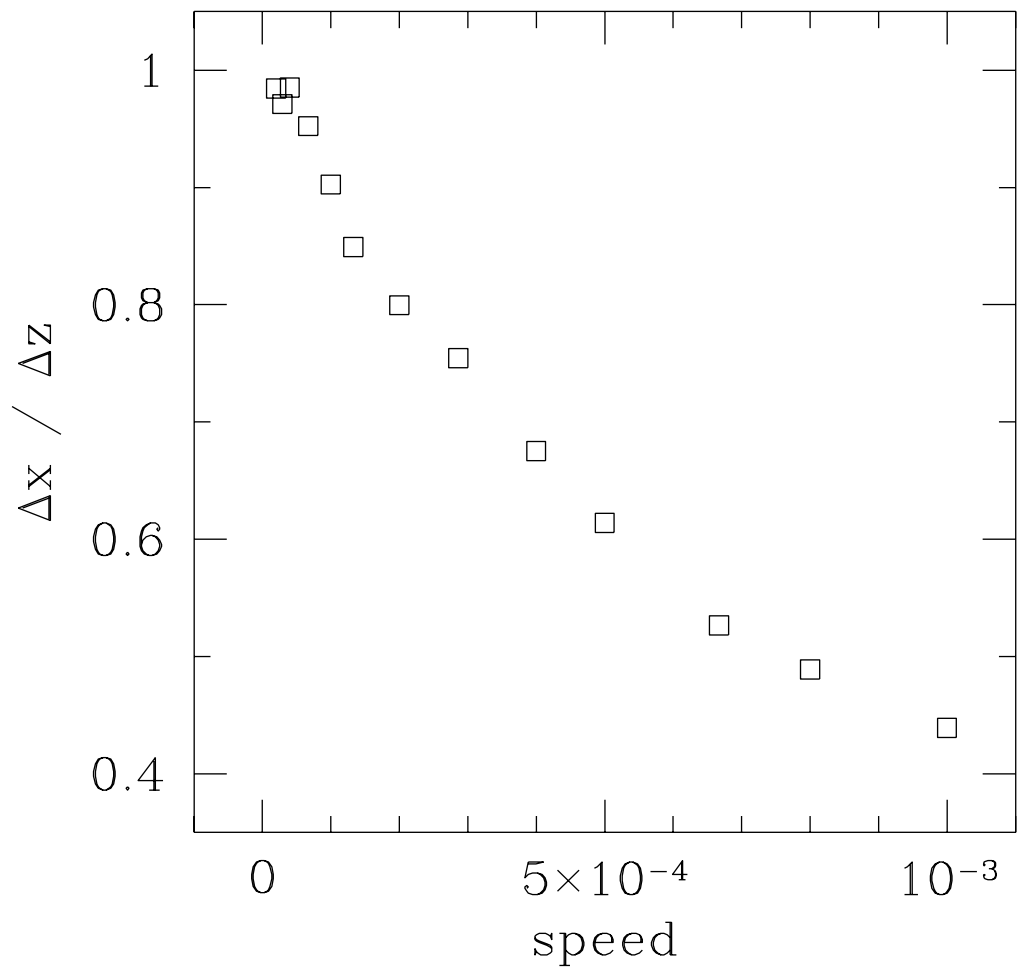,width=6in}
\end{center}
\caption[]{The ratio of the horizontal to vertical radius
of gyration, plotted as a function of $v$}
\label{fig:shape}
\end{figure}

\begin{figure}[p]
\begin{center}
\ 
\psfig{file=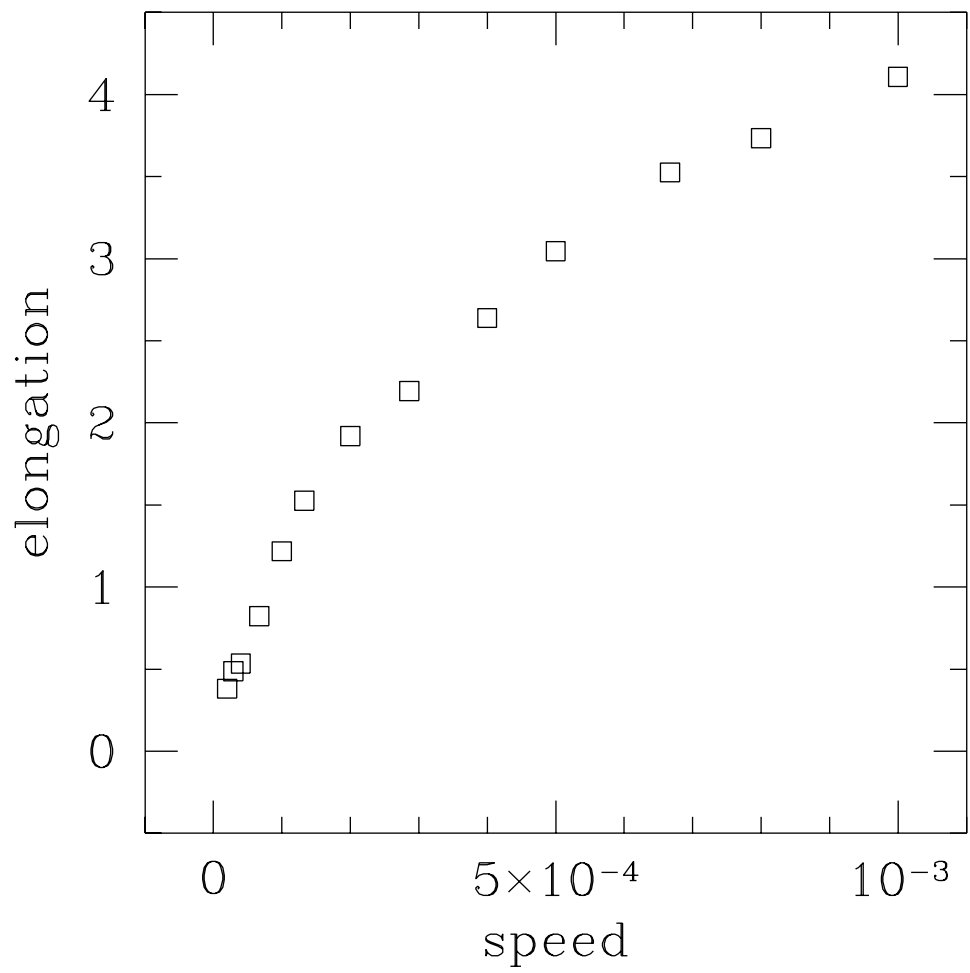,width=6in}
\end{center}
\caption[]{The distance between the head of the chain
and its center of mass as a function of $v$}
\label{fig:elongation}
\end{figure}

In \fig{shape} we plot the ratio of the radius of gyrations
in the horizontal and vertical directions. It decreases to
about half of its equilibrium value. This shows that even
at the highest speeds we were able to consider, there are
still large fluctuations in the shape of the molecule. The
stick-like region of the molecule on average is still not
the entire chain length. This is confirmed further in
\fig{elongation}. Here we plot the distance between the
head of the chain and its center of mass, in the vertical
direction. At the highest speeds this distance is about
$4$.

\begin{figure}[p]
\begin{center}
\ 
\psfig{file=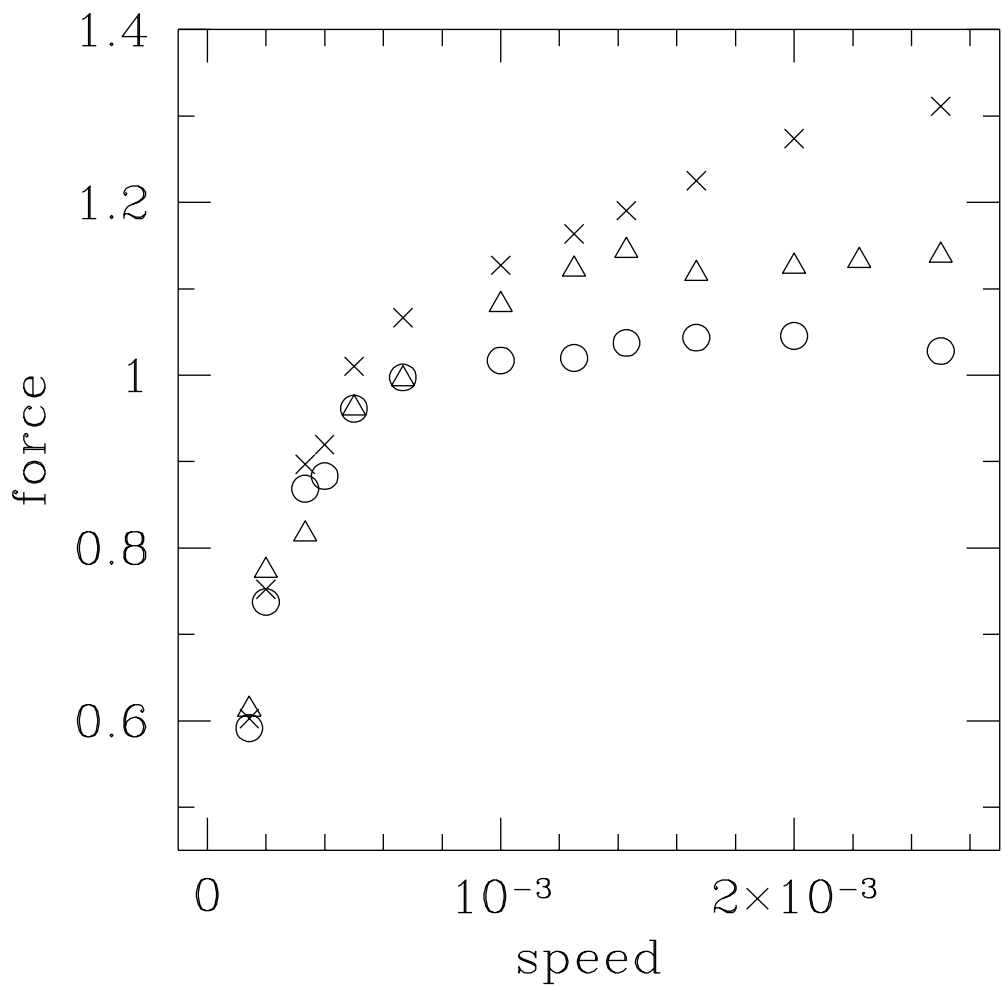,width=6in}
\end{center}
\caption[]{The same as \fig{force} but using long range
moves.}
\label{fig:longrange}
\end{figure}

The force versus velocity for the long range move case 
is shown in \fig{longrange}.
Because the algorithm allows for faster diffusion of kinks,
one can go out to higher velocities. The results however
look quite similar.

\begin{figure}[p]
\begin{center}
\ 
\psfig{file=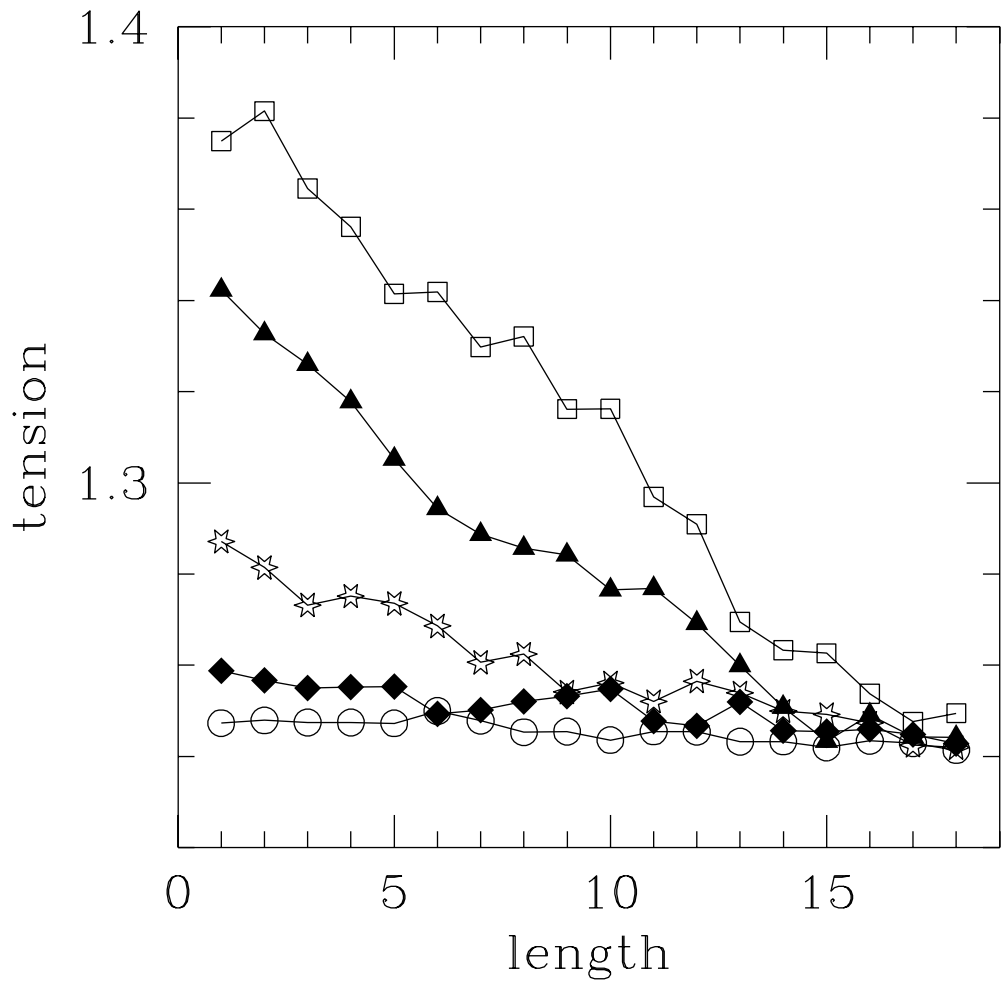,width=6in}
\end{center}
\caption[]{The tension of the chain as a function of
arclength for different speeds. Different symbols
represent speeds as follows. Open squares: $v = 0.001$,
filled triangles:  $v = 0.000667$, stars: $v = 0.0004$,
filled diamonds $v = 0.000133$, open circles: $v = 0.00002$}
\label{fig:tension}
\end{figure}

The tension along the backbone of the chain was measured
for the short range algorithm. It was defined as follows.
We consider two adjacent bonds at some position
along the chain, of which there are
$4\times 4 = 16$ possible conformations. The tension is
viewed as the force along the primitive path of the
chain that stops it from collapsing. We consider the length
of the primitive path for this two bond segment.
When the two bonds have formed a kink, the primitive
path $s$ for this conformation is zero.
When the two bonds are straight, or form a right angle,
the primitive path is 2. Therefore the average length
of the primitive path $\la s\ra$ is related to
the tension $T$ by summing over all conformations
have the same primitive path. For example if the primitive
path is in the shape of a right angle, the chain
can either be in four kink configurations or be a
right angle. There are five conformations altogether
associated with a single primitive path
that are summed over according to the formula
\begin{equation}
\la s\ra = {\sum s e^{-Ts} \over \sum e^{-Ts}}
\label{eq:tension}
\end{equation}
where again we have set the temperature to one.
This simply relates $\la s\ra$ to the local kink 
density. The lower the density, the higher the
tension. Note also that with no applied force
there is still a finite tension. This is to be expected
and it is the entropic ghost force that stops the
chain from collapsing in its 
tube\cite{MDSFE86:TheoryPolymer}.
By measuring $\la s\ra$ one can find the
tension through \eqn{tension}. The tension
as a function of arclength is plotted in \fig{tension}
for different velocities.
One sees a linear decrease in the tension as a function
of arclength, and an overall increase in its magnitude
as $v$ increases. This is in accord with theoretical
predictions\cite{MRAAetal:SlippageRubber} and
is due to the local Rouse friction of the chain in the
tube. The data for the long range moves shows
a much smaller slope and therefore a much lower
friction coefficient. This is reasonable as the
long range moves increase kink mobility.

\begin{figure}[p]
\begin{center}
\ 
\psfig{file=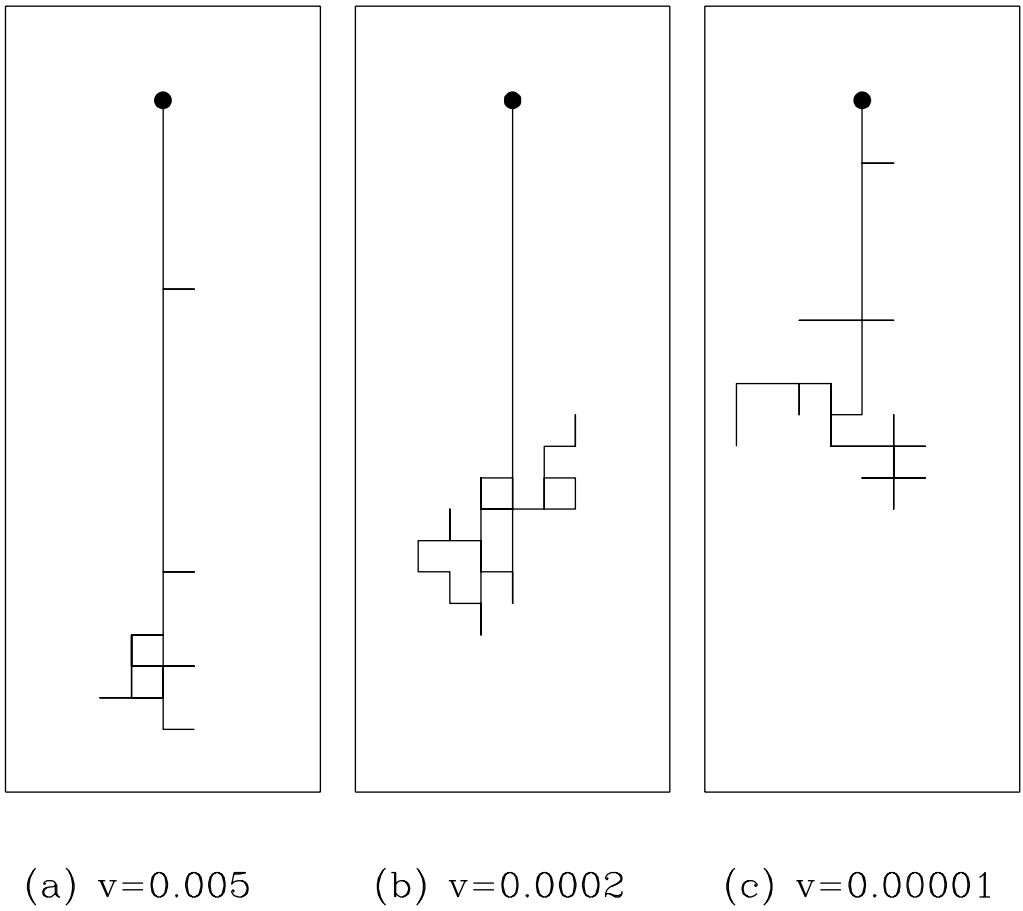,width=6in}
\end{center}
\caption[]{Three conformations are shown for chains
of length 50 at different speeds}
\label{fig:conformations}
\end{figure}

Finally in \fig{conformations} we present pictures of a chain of length
50 moving at three different velocities. For chains
of this length, it is clear that the 
description of the system in terms of a plume and
and stick-like region is correct.

\section{Conclusion}
\label{sec:conclusion}

The measurement of forces in polymer simulations is difficult because
fluctuations are large compared to the size of the mean. In the cases
considered here, we are trying to obtain forces that are the same or
smaller than $k_B T$ divided by a bond length. Because of this,
efficient simulation algorithms are necessary. Models that are
continuous in space and time, such as molecular dynamics, give the
force directly. However they are very slow in comparison with the ones
considered here. They are not viable techniques with 
present computers. Using the
lattice algorithms described in \se{models} we were able to devise a
method that gives an estimate for the force on the end of the chain.
Variants of this method give upper and lower bounds for this force. The
methods break down for sufficiently high pulling speeds, as the head of
the chain cannot keep up with the end of the spring. 
Still results were obtained
for a wide range of velocities giving useful information for the force,
tension, and chain conformation.

The results confirm the picture 
\cite{MRAAetal:SlippageRubber} 
that there is a plateau region of the
force as a function of velocity and the value of the force is close to
$k_B T$ over the entanglement spacing. Because the length of chain we
were able to consider was not very large, but still respectably long by
experimental standards, there are non-negligible fluctuations in
configurations. The length of the plume and the stick-like regions of
the chain varied considerably between different configurations as a
consequence of breathing modes of the chain in the tube. Furthermore,
the plateau region of force versus velocity is not very large
for length $20$ chains. At pulling speeds in the plateau region, the
friction term contributes giving a finite linear slope that can
be discerned from \fig{force}. The long range move case
shows this effect to a lesser extent. These problems could 
certainly be improved by investing more computer time and running
longer chains.

However it is probably more interesting to
exploit the methods we have
developed to consider forces in other problems. One such problem would
be to consider forces of tethered chains at interfaces instead of the
rather artificial problem of a single chain being pulled in steady
state through a gel%
\cite{ERPGdG:RubberRubber,%
HRBCYHetal:InterplayIntermolecular,%
KPOTCBM:MolecularVelcro,%
HRB:EffectsChain}.
In the former case there is the added ingredient
that chains are interacting with a surface that is likely to lead to
interesting physics.

\section*{Acknowledgment}

We would like to thank H.R. Brown for valuable discussions.
This work was supported by NSF Grant DMR-9112767.

\end{document}